\documentclass[aps,pra,twocolumn,showpacs,superscriptaddress,10pt]{revtex4}
\usepackage{graphicx}
\usepackage{subfigure}
\usepackage{amsmath}
\usepackage{amsfonts}
\usepackage{amssymb}
\usepackage{amsthm}
\usepackage{times}
\usepackage[colorlinks,citecolor=blue,linkcolor=blue]{hyperref}
\usepackage{color}

\begin{document}

\title{Signature of non-Markovianity in time-resolved energy transfer}

\newcommand{\Fudan}{\affiliation{State Key Laboratory of Surface
Physics and Department of Physics, Fudan University, Shanghai 200433,
China}}
\newcommand{\Nanjing}{\affiliation{Collaborative Innovation Center of Advanced Microstructures, Fudan University, Shanghai 200433, China}}

\author{Junjie Liu}
\email{jj\_liu@fudan.edu.cn}
\Fudan

\author{Hui Xu}
\Fudan

\author{Chang-qin Wu}
\Fudan
\Nanjing

\date{\today}

\begin{abstract}
We explore signatures of the non-Markovianity in the time-resolved energy transfer processes for quantum open systems. Focusing on typical systems such as the exact solvable damped Jaynes-Cummings model and the general spin-boson model, we establish quantitative links between the time-resolved energy current and the symmetric logarithmic derivative quantum Fisher information (SLD-QFI) flow, one of measures quantifying the non-Markovianity, within the framework of non-Markovian master equations in time-local forms. We find in the damped Jaynes-Cummings model that the SLD-QFI backflow from the reservoir to the system always correlates with an energy backflow, thus we can directly witness the non-Markovianity from the dynamics of the energy current. In the spin-boson model, the relation is built on the rotating-wave approximation, calibrated against exact numerical results, and proven reliable in the weak coupling regime. From the relation, we demonstrate that whether the non-Markovianity guarantees the occurrence of an energy backflow depends on the bath spectral function. For the Ohmic and sub-Ohmic cases, we show that no energy backflow occurs and the energy current always flow out of the system even in the non-Markovian regime. While in the super-Ohmic case, we observe that the non-Markovian dynamics can induce an energy backflow.

\end{abstract}


\maketitle

\section{Introduction}
The unavoidable interaction of the quantum system with its environment leads to dissipation and decoherence processes which strongly modify the dynamics of the system. In the well-established framework of quantum open systems \cite{Breuer.07.NULL,Weiss.12.NULL}, the dynamics of the system is totally determined by the reduced density matrix and the environment-induced effects manifest themselves in the nonunitary time evolution of the reduced density matrix. Under the condition of short environmental correlation times, one can formulate the evolution of the reduced density matrix by means of a dynamical semigroup with a corresponding time-independent generator in Lindblad form \cite{Lindblad.76.CMP,Breuer.07.NULL}, a so-called Markovian dynamics is thus defined. If the dynamics of an open system substantially deviates from that of a dynamical semigroup, one encounter a non-Markovian process. Due to an important role of non-Markovian effects in many realistic experimental scenarios \cite{Li.11.PRA,Cialdi.11.PRA,Liu.11.NP,Smirne.11.PRA,Barreiro.11.N,Gessner.14.NP,Tang.15.O}, a series of measures are proposed to quantify the degree of quantum non-Markovianity \cite{Wolf.08.PRL,Breuer.09.PRL,Rivas.10.PRL,Lu.10.PRA,Chruscinski.11.PRA,Luo.12.PRA,Chruscinski.14.PRL}. In despite of distinct forms of those measures, non-Markovian open quantum systems can be basically characterized via their capability to gain back information previously lost due to decoherence \cite{Breuer.16.RMP}.

Besides the reduced dynamics of open systems, the exchange of energy between the open system and its environment also draw a great deal of attention during recent years (see \cite{Segal.05.PRL,Ren.10.PRL,Boudjada.14.JPCA,Wang.15.SR,Liu.16.NULL,Carrega.16.PRL} and references therein). Noting the energy transfer properties are closely related to the dynamical characteristics of open systems, a natural question about the signature of quantum non-Markovianity in energy transfer processes arises. Recently, a qualitative connection between an energy backflow obtained in the full counting statistic formalism and the non-Markovianity measured by the trace distance \cite{Breuer.09.PRL} is built in the spin-boson model \cite{Guarnieri.16.PRA}. However, an independent qualitative study \cite{Schmidt.16.PRA} demonstrated that the backflow of the trace distance did not necessarily correlate with the energy backflow in the spin-boson model based on the energy current obtained from the definition of work via the power operator \cite{Solinas.13.PRB,Schmidt.15.PRB}. Such a behavior is also found in the quantum Brownian motion by utilizing the Gaussian interferometric power as a non-Markovianity measure \cite{Guarnieri.16.PRAa}. Nevertheless, a quantitative study addressing the role of non-Markovianity in the energy transfer process is still absent, explicit and analytical connections between the non-Markovianity and dynamics of energy transfer are called for in order to make conclusive statements.

In this paper, among the different criteria and measures that quantify quantum non-Markovianity \cite{Wolf.08.PRL,Breuer.09.PRL,Rivas.10.PRL,Lu.10.PRA,Chruscinski.11.PRA,Luo.12.PRA,Chruscinski.14.PRL},   we focus on the one which utilizes the symmetric logarithmic derivative quantum Fisher information (SLD-QFI) flow as the fingerprint \cite{Lu.10.PRA}. As a promising measure, the SLD-QFI flow coincides with the time derivative of the trace distance in exact solvable models \cite{Lu.10.PRA,Haikka.12.PRA}. However, the SLD-QFI scheme only needs an optimal input state instead of optimal state pairs in obtaining the trace distance \cite{He.11.PRA}. Furthermore, the SLD-QFI is directly related to the imaginary part of the dynamical susceptibility \cite{Hauke.16.NP}, the latter is shown to determine the energy transfer properties in the spin-boson model \cite{Liu.16.NULL}, such a connection implies quantitative links between the non-Markovianity and the energy current from the SLD-QFI point of view.

We concentrate our attention to the exact solvable damped Jaynes-Cummings model and the spin-boson model, both described within non-Markovian master equations in time-local forms. We obtain explicit relations between SLD-QFI flow and time-resolved energy current for these two models. Noting the models we considered are paradigms in the theoretical studies of dynamics of quantum open systems \cite{Breuer.07.NULL,Weiss.12.NULL}, thus our results possess good adaptability.

From the so-obtained relationships, we find in the damped Jaynes-Cummings model that the SLD-QFI backflow from the reservoir to the system always correlates with the energy backflow, thus we can witness quantum non-Markovianity directly from the dynamics of the energy current. In the spin-boson model, the relation is built on the rotating-wave-approximation (RWA), benchmarks against the quasiadiabatic propagator path integral (QuAPI) shows that our theory is reliable in the weak coupling regime. We demonstrate that whether the non-Markovianity guarantees the occurrence of an energy backflow depends on the bath spectral function. For the Ohmic as well as sub-Ohmic cases, no energy backflow occurs and the energy current always flow out of the system even in the non-Markovian regime. For the super-Ohmic case, we observe that the non-Markovian dynamics can have an energy backflow. Compared with previous studies, our deductions are completely
analytical and hence the results are more convincing. Moreover, our scheme is helpful for an experimental determination of SLD-QFI as well as the non-Markovianity since the energy current is an experimentally measurable quantity.

The article is organized as follows. In Sec. \ref{sec:1}, we introduce interaction models, the corresponding time-local master equations and their solutions are also recalled. In Sec. \ref{sec:2}, we first briefly review the SLD-QFI and its properties, then derive explicit relations between SLD-QFI flow and transient energy current for models we considered. Finally, in Sec. \ref{sec:3}, we discuss the results in detail and reveal the signature of non-Markovianity in energy transfer processes. Conclusions are
drawn in Sec. \ref{sec:4}.

\section{Model and master equations}\label{sec:1}
General quantum open systems are governed by the following Hamiltonian
\begin{equation}
H~=~H_s+H_I+H_B,
\end{equation}
where $H_s$ and $H_B$ are the Hamiltonians of system and environment, respectively, and $H_I$ denotes a system-environment interaction Hamiltonian. Given a factorized initial system-environment state, the evolution of the reduced density matrix $\rho_s$ can be described by the time-local master equations \cite{Breuer.07.NULL,Chruscinski.10.PRL}
\begin{equation}\label{eq:tlme}
\frac{d}{d t}\rho_s(t)~=~\mathcal{K}(t)\rho_s(t),
\end{equation}
in order to preserve the Hermiticity and trace of the density matrix, the time-dependent generator $\mathcal{K}$ must be of the form in the Schr\"{o}dinger picture \cite{Gorini.76.JMP,Breuer.04.PRA,Breuer.09.PRL,Chruscinski.10.PRL,Breuer.16.RMP}
\begin{eqnarray}\label{eq:gene}
\mathcal{K}(t)\rho_s &=& -i[\tilde{H}_s(t),\rho_s]+\sum_i\gamma_i(t)\left[A_i(t)\rho_s A_i^{\dagger}(t)\right.\nonumber\\
&&-\left.\frac{1}{2}\{A_i^{\dagger}(t)A_i(t),\rho_s\}\right],
\end{eqnarray}
where $\tilde{H}_s(t)=H_s+H_{LS}(t)$ with $H_{LS}(t)$ the time-dependent Lamb shift Hamiltonian, $\gamma_i(t)$ and $A_i(t)$ denote the relaxation rates and Lindblad operators, respectively, $\{P,Q\}$ is the anticommutator for arbitrary operators $P$ and $Q$. Master equations in such time-local forms can be derived by employing the time-convolutionless projection operator technique \cite{Breuer.07.NULL}. If the Hamiltonian $\tilde{H}_s(t)$, the relaxation rates $\gamma_i(t)$ and Lindblad operators $A_i(t)$ are time independent, and all $\gamma_i$ are positive, the above master equation reduces to the well-known Lindblad equation describing the conventional Markovian process \cite{Lindblad.76.CMP}. However, in the time-dependent case $\gamma_i(t)$ can become temporarily negative without violating complete positivity, thus leading to the non-Markovian dynamics. In this study, we focus on two specific interaction models whose master equations have the time-local forms and for which signatures of non-Markovianity in the energy transfer process can be revealed explicitly.

\subsection{The Damped Jaynes-Cummings model}
The first one is the damped Jaynes-Cummings model \cite{Breuer.07.NULL,Lu.10.PRA,Zeng.11.PRA,Breuer.16.RMP}. The system Hamiltonian $H_s$ and the environmental Hamiltonian $H_B$ are given by
\begin{equation}\label{eq:free_H}
H_s~=~\omega_0\sigma_+\sigma_-,~~~H_B~=~\sum_k\omega_kb_k^{\dagger}b_k,
\end{equation}
where $\omega_0$ is the energy spacing between ground state $|g\rangle$ and excited state $|e\rangle$ of the system, $\sigma_-=|g\rangle\langle e|$ and $\sigma_+=|e\rangle\langle g|$ are the lowering and raising operators of the system, $b_k$ and $b_k^{\dagger}$ are annihilation and creation operators for the bosonic reservoir mode labeled by $k$ with frequency $\omega_k$. The interaction term is taken to be
\begin{equation}
H_I~=~\sum_k(\sigma_+g_kb_k+\sigma_-g_k^{\ast}b_k^{\dagger})
\end{equation}
with coupling constants $g_k$.

Due to the conservation of the number of excitations, this model can be exactly solvable. For the sake of completeness, we briefly recall the exact results. In the interaction picture, the exact master equation for $\rho(t)$ has the time-local form of Eq. (\ref{eq:tlme})\cite{Breuer.07.NULL}
\begin{eqnarray}\label{eq:tlme_JC}
\frac{d}{d t}\rho_s(t) &=& -\frac{i}{2}S(t)[\sigma_+\sigma_-,\rho_s(t)]+\gamma(t)\left[\sigma_-\rho_s(t)\sigma_+\right.\nonumber\\
&&-\left.\frac{1}{2}\{\sigma_+\sigma_-,\rho_s(t)\}\right],
\end{eqnarray}
where the time-dependent Lamb shift $H_{LS}(t)=S(t)\sigma_+\sigma_-$ with $S(t)=-2\mathrm{Im}\left[\dot{G}(t)/G(t)\right]$ and the relaxation rate $\gamma(t)=-2\mathrm{Re}\left[\dot{G}(t)/G(t)\right]$, the function $G(t)$ is totally determined as the solution of the integral equation
\begin{equation}
\frac{d}{d t}G(t)~=~-\int_0^td\tau f(t-\tau)G(\tau)
\end{equation}
with $f(t-\tau)=\int d\omega J(\omega)\mathrm{exp}[i(\omega_0-\omega)(t-\tau)]$ the inverse Fourier transform of the bath spectral density $J(\omega)$ and the initial condition $G(0)=1$.

The model was frequently used to describe the atom-cavity system \cite{Breuer.07.NULL}, so we usually choose the Lorentzian spectral function
\begin{equation}
J(\omega)~=~\frac{1}{2\pi}\frac{\gamma_0\lambda^2}{(\omega_0-\omega)^2+\lambda^2}
\end{equation}
with $\lambda$ the width and $\gamma_0$ the strength of the system-environment coupling. Then the function $G(t)$ takes the form
\begin{equation}
G(t)~=~e^{-\lambda t/2}\left[\cosh\left(\frac{dt}{2}\right)+\frac{\lambda}{d}\sinh\left(\frac{dt}{2}\right)\right],
\end{equation}
where $d=\sqrt{\lambda^2-2\gamma_0\lambda}$. Since $G(t)$ is real, we will obtain a vanishing Lamb shift $S(t)=0$, and the relaxation rate $\gamma(t)$ reads
\begin{equation}\label{eq:JC_gamma}
\gamma(t)~=~\frac{2\gamma_0\lambda\sinh(dt/2)}{d\cosh(dt/2)+\lambda\sinh(dt/2)}.
\end{equation}
For weak couplings, corresponding to $\gamma_0<\lambda/2$, Eq. (\ref{eq:JC_gamma}) is always positive such that the dynamics is Markovian. In the limit of $\gamma_0\ll\lambda/2$ the rate $\gamma(t)$ becomes time independent, then the master equation Eq. (\ref{eq:tlme_JC}) is of the Lindblad form and leads to a dynamical semigroup. However, in the strong coupling regime, namely, for $\gamma_0>\lambda/2$, $\gamma(t)$ will take negative values, implying that a non-Markovian dynamics emerges.

Considering the general initial condition for the reduced density matrix
\begin{equation}\label{eq:rho_initial}
\rho_s(0)~=~\frac{1}{2}\left(
\begin{array}{cc}
\cos\eta+1 & \sin\eta\\
\sin\eta & 1-\cos\eta
\end{array}
\right)
\end{equation}
with $\eta$ a real parameter. We can obtain an explicit form for $\rho_s(t)$ from the master equation Eq. (\ref{eq:tlme_JC}) as follows
\begin{equation}\label{eq:rho_JC}
\rho_s(t)~=~\frac{1}{2}\left(
\begin{array}{cc}
(\cos\eta+1)G^2(t) & \sin\eta G(t)\\
\sin\eta G(t) & 2-(\cos\eta+1)G^2(t)
\end{array}
\right).
\end{equation}

\subsection{The spin-boson model}
The second model is the spin-boson model \cite{Leggett.87.RMP,Weiss.12.NULL}. As a minimal prototype in studying nontrivial effects induced by the environment, it can describe many physical realizations, such as electron-transfer reactions \cite{Merkli.13.JMC}, biomolecules \cite{Garg.85.JCP}, superconducting circuits \cite{Weiss.12.NULL}, tunneling light particles in metals \cite{Fukai.05.NULL}, anomalous low temperature thermal properties in glasses \cite{Anderson.72.PM}, to mention just a few. The environment is still taken to be a bosonic bath and the corresponding Hamiltonian is again given by Eq. (\ref{eq:free_H}), but the system Hamiltonian and the interaction term are replaced by
\begin{equation}\label{eq:SBM}
H_s~=~\frac{\omega_0}{2}\sigma_z,~~~H_I~=~\sigma_x\sum_k(g_kb_k+g_k^{\ast}b_k^{\dagger}),
\end{equation}
where $\sigma_{z,x}$ denote the usual Pauli matrices, other notations remain the same meanings with the damped Jaynes-Cummings model.

We assume that the system-environment coupling is weak
and employ the second-order time-convolutionless master
equation \cite{Breuer.07.NULL} within the RWA to describe the evolution of the reduced density matrix. Within the approximation, the Lamb shift is negligible, we have the following Schr\"{o}dinger picture master equation \cite{Haikka.10.PS,Clos.12.PRA,Hao.13.JPA}
\begin{eqnarray}\label{eq:tlme_SBM}
\frac{d}{d t}\rho_s(t) &=& -\frac{i\omega_0}{2}[\sigma_z,\rho_s(t)]+\sum_{m=\pm}\gamma_m(t)\left[\sigma_m\rho_s(t)\sigma_m^{\dagger}\right.\nonumber\\
&&-\left.\frac{1}{2}\{\sigma_m^{\dagger}\sigma_m,\rho_s(t)\}\right],
\end{eqnarray}
where the relaxation rates are determined by
\begin{eqnarray}
\gamma_{\pm}(t) &=& \frac{1}{2}\int\,J(\omega)\left[(1+n_B)\frac{\sin(\omega\pm\omega_0)t}{\omega\pm\omega_0}\right.\nonumber\\
&&\left.+n_B\frac{\sin(\omega\mp\omega_0)t}{\omega\mp\omega_0}\right]d\omega
\end{eqnarray}
with $n_B$ the Bose-Einstein distribution characterized by the temperature $T$.

In order to solve the above master equation, we can write it in terms of
the Bloch vector $\vec{B}(t)=\mathrm{Tr}_s[\vec{\sigma}\rho_s(t)]$ with $\vec{\sigma}=(\sigma_x,\sigma_y,\sigma_z)$ \cite{Breuer.07.NULL}
\begin{equation}\label{eq:BV}
\frac{d}{dt}\vec{B}(t)~=~M(t)\vec{B}(t)+\vec{b}(t),
\end{equation}
the matrix $M(t)$ is given by
\begin{equation}
M(t)~=~\left(
\begin{array}{ccc}
-\frac{1}{2}\gamma_s(t) & -\omega_0 & 0\\
\omega_0 & -\frac{1}{2}\gamma_s(t) & 0\\
0 & 0 & -\gamma_s(t)
\end{array}
\right),
\end{equation}
and the vector $\vec{b}(t)=(0,0,\gamma_d(t))^T$, where we have introduced
\begin{eqnarray}
\gamma_s(t) &=& \gamma_+(t)+\gamma_-(t)=\frac{1}{2}\int_0^tds\cos(\omega_0s)D_1(s),\label{eq:gamma_sum}\\
\gamma_d(t) &=& \gamma_+(t)-\gamma_-(t)=-\frac{1}{2}\int_0^tds\sin(\omega_0s)D(s)
\end{eqnarray}
with $D_1(s)=2\int_0^{\infty}d\omega J(\omega)\coth\frac{\omega}{2T}\cos\omega s$ and $D(s)=2\int_0^{\infty}d\omega J(\omega)\sin\omega s$ the noise and dissipation kernel of the model, respectively.

With the initial condition Eq. (\ref{eq:rho_initial}), we find the solution of Eq. (\ref{eq:BV}) as
\begin{equation}\label{eq:BV_SB}
\vec{B}(t)~=~(e^{-\Lambda}\sin\eta\cos\omega_0t,e^{-\Lambda}\sin\eta\sin\omega_0t,e^{-\Gamma}(\cos\eta+\delta)),
\end{equation}
where $\Gamma(t)=\int_0^t\gamma_s(\tau)d\tau$, $\Lambda(t)=\Gamma(t)/2$ and $\delta(t)=\int_0^te^{\Gamma(\tau)}\gamma_d(\tau)d\tau$. Therefore, the reduced density matrix can be expressed as
\begin{equation}
\rho_s(t)=\frac{1}{2}(\mathbf{\mathrm{I}}+\vec{B}(t)\cdot\vec{\sigma})
\end{equation}
with $\mathbf{\mathrm{I}}$ the $2\times2$ identity matrix.

\section{Non-Markovianity and time-resolved energy current}\label{sec:2}
In this section, we will build explicit relations between the non-Markovianity and the transient energy current for the above two models.

\subsection{Quantum Fisher information and non-Markovianity}
At first, for the sake of completeness, we briefly review the properties of the SLD-QFI and especially its application in quantifying the non-Markovianity. By applying a phase transformation to the reduced density matrix such that $\rho_s(\theta;t)$ contains a real parameter $\theta$. Then the SLD-QFI is defined as \cite{Helstrom.76.NULL}
\begin{equation}
\mathcal{F}(\theta;t)~=~\mathrm{Tr}[\rho_s(\theta;t)L^2(\theta;t)],
\end{equation}
where $L(\theta;t)$ is the symmetric logarithmic derivative determined by
\begin{equation}
\frac{\partial}{\partial\theta}\rho_s(\theta;t)=\frac{1}{2}\left[L(\theta;t)\rho_s(\theta;t)+\rho_s(\theta;t)L(\theta;t)\right].
\end{equation}
The SLD-QFI plays a vital role in quantum metrology \cite{Giovannetti.06.PRL} and quantum estimation theory \cite{Helstrom.76.NULL}. An equivalence to the fidelity susceptibility even enables the SLD-QFI to analyze quantum phase transitions in the ground state \cite{You.07.PRE,Venuti.07.PRL,Zanardi.07.PRL} and in dissipative systems \cite{Banchi.14.PRE,Marzolino.14.PRA}. Furthermore, multiparticle entanglement can be detected via the SLD-QFI \cite{Hyllus.12.PRA}.

According to the quantum Cram\'{e}r-Rao theorem for the mean-square error $\langle(\delta\theta)^2\rangle$ \cite{Braunstein.94.PRL} of estimation results for the parameter $\theta$ \cite{Helstrom.76.NULL}
\begin{equation}
\langle(\delta\theta)^2\rangle\geqslant\frac{1}{N\mathcal{F}(\theta;t)}
\end{equation}
with $N$ the times of independent measurements, we know that the SLD-QFI imposes an upper bound on the precision of parameter estimation. A larger SLD-QFI implies that the parameter $\theta$ can be estimated with higher precision.

Here we consider the parameter $\theta$ introduced onto the system through the following interferometer \cite{Hyllus.10.PRA}
\begin{equation}
\rho_s(\theta;t)~=~e^{i\theta J_n}\rho_s(t)e^{-i\theta J_n},
\end{equation}
where $J_n=\vec{n}\cdot\vec{J}$ with $\vec{n}$ an arbitrary direction and $\vec{J}$ the angular momentum. For two-level systems (TLSs), we have $\vec{J}=\vec{\sigma}/2$. Under such a transformation, the resulting SLD-QFI is independent of the parameter $\theta$ and takes a simple form in TLSs \cite{Suzuki.16.PRA}
\begin{equation}\label{eq:QFI_general}
\mathcal{F}(\theta;t)=\mathcal{F}(t)=\left|\vec{n}\times \vec{B}(t)\right|^2
\end{equation}
with $\vec{B}(t)$ the Bloch vector from the reduced density matrix. Therefore, the results we obtain below is independent of extra parameters and just manifests intrinsic properties of the systems we considered.

Since only the maximum SLD-QFI is of physical significance, we should firstly choose an optimal direction $\vec{n}_o$ in Eq. (\ref{eq:QFI_general}) which corresponds to an optimal experimental setup (see the details in the Appendix \ref{sec:ap_od}), then adopt an optimal initial spin state to further maximize the SLD-QFI. By doing so, the maximum SLD-QFI of TLSs reads
\begin{equation}\label{eq:QFI_optimal}
\mathcal{F}_M(t)=\left|\vec{B}_M(t)\right|^2
\end{equation}
with $\vec{B}_M(t)$ the Bloch vector Eq. (\ref{eq:BV_SB}) obtained from the optimal initial spin state. Since the initial condition Eq. (\ref{eq:rho_initial}) corresponds to a pure state, then the initial maximum SLD-QFI $\mathcal{F}_M(0)$ is equal to 1 according to the property of the Bloch sphere. If the dynamics is Markovian, the evolution of the open system will be governed by a completely positive and trace-preserving map (or a dynamical semigroup) and the maximum SLD-QFI $\mathcal{F}_M(t)$ of the system will monotonically decreases from 1 \cite{Fujiwara.01.PRA}, meaning that information continuously flow out of the system towards the environment. Analogously, a temporary increase of the maximum SLD-QFI $\mathcal{F}_M(t)$ can be ascribed to a backflow of information from the environment to the system again. Non-Markovian quantum dynamics are accordingly defined as those which show a nonmonotonic behavior of the maximum SLD-QFI $\mathcal{F}_M(t)$, similar to the situation of the trace distance \cite{Breuer.09.PRL}. Then we can introduce the SLD-QFI flow
\begin{equation}\label{eq:IQ}
\mathcal{I}_Q(t)~=~\frac{d}{dt}\mathcal{F}_M(t)
\end{equation}
to quantify the non-Markovianity \cite{Lu.10.PRA}. Once the SLD-QFI flow becomes positive at some time interval, it signifies the emergence of the non-Markovian dynamics.

\subsection{Information flow and energy current}
In this subsection, we will establish explicit relations between the SLD-QFI flow [Eq. (\ref{eq:IQ})] and the time-resolved energy current. Since we focus on the dynamics of the open system and note that the maximum SLD-QFI is totally determined by the reduced density matrix, correspondingly, we consider the following definition for the time-resolved energy current
\begin{equation}\label{eq:IE}
\mathcal{I}_E~=~\frac{d}{dt}\langle H_s\rangle=\mathrm{Tr}[\dot{\rho}_s(t)H_s],
\end{equation}
where $H_s$ is the system Hamiltonian. If $\mathcal{I}_E>0$, meaning that we have a backflow of energy from the environment to the system.

For the damped Jaynes-Cummings model, we have already known that \cite{Lu.10.PRA}
\begin{equation}\label{eq:IQ_JC}
\mathcal{I}_Q(t)~=~2G(t)\dot{G}(t)
\end{equation}
with the optimal initial condition corresponds to $\sin\eta=1$ in Eq. (\ref{eq:rho_initial}). Under the same initial condition, the reduced density matrix Eq. (\ref{eq:rho_JC}) reads
\begin{equation}\label{eq:rho_JC_O}
\rho_s(t)~=~\frac{1}{2}\left(
\begin{array}{cc}
G^2(t) & G(t)\\
G(t) & 2-G^2(t)
\end{array}
\right).
\end{equation}
Then the time-resolved energy current Eq. (\ref{eq:IE}) takes the form
\begin{equation}\label{eq:IE_JC}
\mathcal{I}_E(t)~=~\omega_0G(t)\dot{G}(t).
\end{equation}
Compared the above result with Eq. (\ref{eq:IQ_JC}), we immediately obtain
\begin{equation}\label{eq:relation_JC}
\mathcal{I}_E~=~\frac{\omega_0}{2}\mathcal{I}_Q
\end{equation}
for the damped Jaynes-Cummings model.

For the spin-boson model, we find the optimal initial state that maximizes the SLD-QFI [Eq. (\ref{eq:QFI_optimal})] to be the spin-up state of $\sigma_z$ which corresponds to $\cos\eta=1$ in Eq. (\ref{eq:rho_initial}), therefore the Bloch vector [Eq. (\ref{eq:BV_SB})] becomes $\vec{B}_M(t)=(0,0,e^{-\Gamma}(1+\delta))$ and
\begin{equation}
\mathcal{F}_M(t)~=~e^{-2\Gamma(t)}\left(1+\delta(t)\right)^2.
\end{equation}
Hence the SLD-QFI flow takes the form
\begin{equation}\label{eq:IQ_SBM}
\mathcal{I}_Q(t)~=~2e^{-2\Gamma(t)}\left(1+\delta(t)\right)\left(\dot{\delta}(t)-\dot{\Gamma}(t)(1+\delta(t))\right).
\end{equation}
From the reduced density matrix, we have
\begin{equation}\label{eq:IE_SBM}
\mathcal{I}_E~=~\frac{\omega_0}{2}e^{-\Gamma(t)}\left(\dot{\delta}(t)-\dot{\Gamma}(t)(1+\delta(t))\right),
\end{equation}
Then we can obtain for the spin boson model that
\begin{equation}\label{eq:relation_SBM}
\mathcal{I}_E~=~\frac{\omega_0}{4\langle\sigma_z\rangle}\mathcal{I}_Q,
\end{equation}
where we have used the fact that $B_3$ is just the spin population $\langle\sigma_z\rangle$. 

Eqs. (\ref{eq:relation_JC}) and (\ref{eq:relation_SBM}) are the main results of this study. It is worthwhile to remark that although the definitions Eqs. (\ref{eq:IQ}) and (\ref{eq:IE}) are quite general, a universal link between $\mathcal{I}_Q$ and $\mathcal{I}_E$ is absent since the optimal initial system state in obtaining the maximum SLD-QFI is model-dependent as we saw in the present two models. However, we expect model-dependent connections to exist in open systems beyond paradigms considered here, such as the open multi-level system which desires a further investigation.

Nevertheless, the identification given by Eqs. (\ref{eq:relation_JC}) and (\ref{eq:relation_SBM}) still have several conceptual implications. First, we can study the non-Markovianity from the energy transfer perspective. Experimental advances in the field of energy transfer at nanoscale \cite{Koh.07.PRB,Lefevre.05.PSI,Chavez.14.APLM,Reparaz.14.PSI} even enable us to experimentally determine SLD-QFI as well as non-Markovianity. Second, through rigorous relations we show that the dynamics of open systems inevitably manifests in the energy transfer process.

\section{Discussions}\label{sec:3}
In this section, we will study the signature of the non-Markovianity measured by the SLD-QFI flow in the energy transfer process for the two models by using Eqs. (\ref{eq:relation_JC}) and (\ref{eq:relation_SBM}).

\subsection{The Damped Jaynes-Cummings model}
\begin{figure}[tbh]
  \centering
  \includegraphics[width=1\columnwidth]{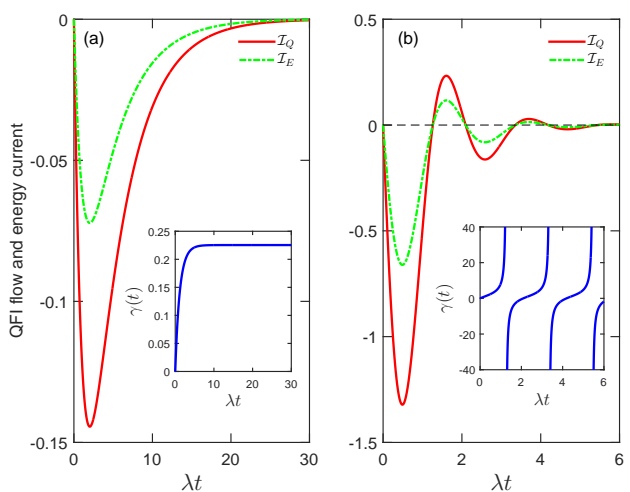}
\caption{(Color online)SLD-QFI flow (solid red line) and transient energy current (dashed-dotted green line) as functions of rescaled time for the damped Jaynes-Cummings model. (a) Weak coupling regime with $\gamma_0=0.2\lambda$; (b) Strong coupling regime with $\gamma_0=5\lambda$. Insets show the relaxation rate $\gamma$ as functions of rescaled time for both regimes. We have chose $\omega_0=1$ as the unit.}
\label{fig:JCtotal}
\end{figure}

We first look at the damped Jaynes-Cummings model. The physical meaning of the relation Eq. (\ref{eq:relation_JC}) is quite apparent. In the weak coupling regime where the dynamics is Markovian, $\mathcal{I}_Q$ and $\mathcal{I}_E$ are always negative as can be seen from Fig. \ref{fig:JCtotal}(a), implying the SLD-QFI and energy are continuously lost during the time evolution of the open system. At sufficient large times when a thermal equilibrium is established between the system and the bosonic bath, they approach zero.

In the strong coupling regime in which the dynamics is non-Markovian, the relaxation rate takes on negative values in some intervals of time as shown in the inset of Fig. \ref{fig:JCtotal}(b). Within those intervals, both $\mathcal{I}_Q$ and $\mathcal{I}_E$ become positive. Thus in this exact solvable model, the SLD-QFI backflow or non-Markovianity always guarantees the occurrence of the energy backflow and we can witness the non-Markovianity directly from the dynamics of the energy exchange.

\subsection{The spin-boson model}
Compared with the above exact solvable system, the interplay between $\mathcal{I}_Q$ and $\mathcal{I}_E$ in the spin-boson model is complicated due to the evolution of the spin population as illustrated by Eq. (\ref{eq:relation_SBM}).

In the following study, we choose the spectral density as \cite{Weiss.12.NULL}
\begin{equation}\label{eq:bath_spec}
J(\omega)~=~\pi\alpha\omega^s\omega_c^{1-s} e^{-\omega/\omega_c},
\end{equation}
where $\alpha$ is the dimensionless system-bath coupling strength, $\omega_c$ denotes the cut-off frequency. The case $s>1(s<1)$ corresponds to super-Ohmic (sub-Ohmic) dissipation, and $s=1$ represents the important
case of frequency-independent (Ohmic) dissipation. Since we have utilized the RWA in the master equation Eq. (\ref{eq:tlme_SBM}), the results we obtained should be valid in the weak coupling regime \cite{Breuer.07.NULL}.

\begin{figure}[tbh]
  \centering
  \includegraphics[width=1.\columnwidth]{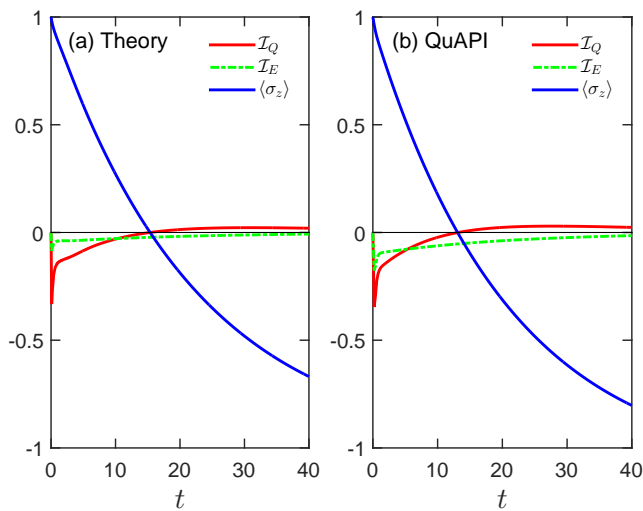}
\caption{(Color online) Evolutions of the SLD-QFI flow $\mathcal{I}_Q$ (solid red line), the transient energy current $\mathcal{I}_E$ (dashed-dotted green line) and the spin population $\langle\sigma_z\rangle$ (solid blue line) in the spin boson model with an Ohmic spectrum. (a) Theory results based on Eqs. (\ref{eq:IQ_SBM}), (\ref{eq:IE_SBM}) and (\ref{eq:BV_SB}); (b) Exact numerical results using QuAPI. We have chose $\omega_0=1$ as the unit, $\omega_c=10$, $\alpha=0.02$ and $T=0.01$.}
\label{fig:SBtotals1}
\end{figure}

We first focus on the Ohmic case. Noting SLD-QFI [Eq. (\ref{eq:QFI_optimal})] and $\mathcal{I}_E$ [Eq. (\ref{eq:IE})] are totally determined by the reduced density matrix $\rho_s(t)$, we can carry out numerical simulations to calculate them for comparisons. Among various numerical methods, we utilize the QuAPI which serves as a benchmark for the spin-boson model \cite{Makri.95.JCP} to obtain exact temporal behaviors for the reduced density matrix. The initial spin state in numerical simulations is taken to be the optimal initial state, namely, the spin-up state of $\sigma_z$. Once the exact evolution of the reduced density matrix $\rho_s(t)$ is obtained, we can get the corresponding numerical results for $\mathcal{I}_Q(t)$ and $\mathcal{I}_E(t)$ according to definitions Eqs. (\ref{eq:IQ}) and (\ref{eq:IE}), respectively.  The results are summarized in Fig. \ref{fig:SBtotals1}. As can be seen, our theory indeed captures essential features of quantities we considered in the weak coupling regime, although minor deviations exist between the theory and exact numerical results.
From the figure, we observe that the SLD-QFI flow $\mathcal{I}_Q$ is negative only at short time scales, after a certain time, a backflow will occur and the non-Markovian dynamics emerges. In contrast, the transient energy current $\mathcal{I}_E$ is always negative even at the non-Markovian regime due to the sign change of the spin population. Thus in this case the non-Markovianity does not guarantees the occurrence of the energy backflow, in accordance with the finding in Ref.\cite{Schmidt.16.PRA}.

\begin{figure}[tbh]
  \centering
  \includegraphics[width=1\columnwidth]{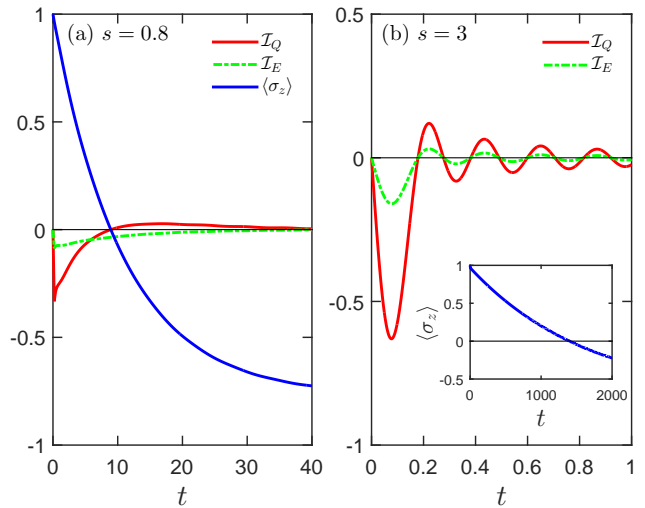}
\caption{(Color online) Theoretical predictions for the SLD-QFI flow $\mathcal{I}_Q$ (solid red line, Eq. (\ref{eq:IQ_SBM})), the transient energy current $\mathcal{I}_E$ (dashed-dotted green line, Eq. (\ref{eq:IE_SBM})) and the spin population $\langle\sigma_z\rangle$ (solid blue line, Eq. (\ref{eq:BV_SB})) in the spin boson model with non-Ohmic spectrums: (a) a sub-Ohmic spectrum with $s=0.8$, (b) a super-Ohmic spectrum with $s=3$. We have chose $\omega_0=1$ as the unit, $\omega_c=10$, $\alpha=0.02$ and $T=0.5$ for both cases.}
\label{fig:SBtotals3}
\end{figure}

We then turn to the non-Ohmic case. For sub-Ohmic spectrums, we take $s=0.8$ as an illustration, other values with $s<1$ depict similar behaviors. For super-Ohmic spectrums, by taking the possible experimental relevance into account, we choose $s=3$ since it can apply to a defect tunneling in the b.c.c. materials Nb and Fe with coupling to acoustic phonons \cite{Goerlich.89.EL}. Since our theory is correct in the weak coupling regime, we only present theoretical results in Fig. \ref{fig:SBtotals3} for simplicity.

As can be seen from Fig. \ref{fig:SBtotals3}(a), the evolution behaviors of $\mathcal{I}_Q$ and $\mathcal{I}_E$ in the sub-Ohmic case are quite similar to those in the Ohmic case, thus we can conclude that for $s\leq1$ there is no energy backflow even in the non-Markovian regime. While for the super-Ohmic case as shown in Fig. \ref{fig:SBtotals3}(b), both $\mathcal{I}_Q$ and $\mathcal{I}_E$ depict oscillating behaviors as a function of time. For positive spin populations, those oscillations ensure that the SLD-QFI backflow always accompanies an energy backflow. Noting the time from which the spin population becomes negative is quite large compared with the time scale $\omega_0^{-1}$ of the system as shown in the inset, thus we can observe the energy backflow in the non-Markovian regime for a long time.

\section{Summary}\label{sec:4}
We have studied signatures of the non-Markovianity in the transient energy current for typical quantum open systems. By utilizing the SLD-QFI flow, one of measures quantifying the non-Markovianity, we establish explicit relations between the non-Markovianity and transient energy current for the damped Jaynes-Cummings model and the spin-boson model, both described within non-Markonian master equations in time-local forms.

In the exact solvable damped Jaynes-Cummings model, the relation between SLD-QFI flow and the transient energy current clearly reveals that the SLD-QFI backflow always correlates with an energy backflow, thus we can witness the non-Markovianity directly from the dynamics of energy exchange. By contrast, in the spin-boson model, we focus on the weak coupling regime where the theory is valid as confirmed by the quasiadiabatic propagator path integral and demonstrate that whether the non-Markovianity guarantees the occurrence of an energy backflow depends on the bath spectral function. In the Ohmic as well as sub-Ohmic case, there is simply no energy backflow occurs and the energy current always flow out of the system even in the non-Markovian regime. While in the super-Ohmic case, we observe that an energy backflow occurs whenever a non-Markovian dynamics emerges during a long time interval.

In perspective, it would be interesting for us to go beyond the rotating-wave-approximation and investigate the strong coupling regime of the spin-boson model where non-Markovian effects are significant. The generality of the present result also needs a thorough investigation.

\begin{acknowledgments}
Support from the National Nature Science Foundation of China with Grant No. 11574050 is gratefully acknowledged.
\end{acknowledgments}

\appendix
\section{Optimal directions}\label{sec:ap_od}
With the diagonal form of the reduced density matrix $\rho_s=\sum_{i}p_i|i\rangle\langle i|$, the optimal direction $\vec{n}_o$ is determined by the eigenvector of a matrix $C$ with the maximal eigenvalue, the entries of the matrix $C$ reads \cite{Hyllus.10.PRA,Hao.13.JPA}
\begin{eqnarray}
[C]_{kl} &=& \sum_{i\neq j}\frac{(p_i-p_j)^2}{p_i+p_j}\left[\langle i|J_k|j\rangle\langle j|J_l|i\rangle\right.\nonumber\\
&&+\left.\langle i|J_l|j\rangle\langle j|J_k|i\rangle\right],
\end{eqnarray}
where $J_k$ is the $k$th component of the angular momentum vector $\vec{J}$.

Noting for the TLS, $\rho_s(t)=(\mathbf{\mathrm{I}}+\vec{B}\cdot\vec{\sigma})/2$ with $\vec{B}=(B_1,B_2,B_3)$, thus the eigenvectors of $\rho_s$ take the form
\begin{eqnarray}
|1\rangle &=& \cos\frac{\psi}{2}e^{-i\xi/2}|e\rangle+\sin\frac{\psi}{2}e^{i\xi/2}|g\rangle,\\
|2\rangle &=& -\sin\frac{\psi}{2}e^{-i\xi/2}|e\rangle+\cos\frac{\psi}{2}e^{i\xi/2}|g\rangle
\end{eqnarray}
with $\tan\xi=B_2/B_1$, $\tan\psi=\sqrt{B_1^2+B_2^2}/B_3$ and $|g\rangle,|e\rangle$ the eigenvectors of $\sigma_z$. The corresponding eigenvalues are given by
\begin{eqnarray}
p_1 &=& \frac{1}{2}+\frac{1}{2}\sqrt{B_3^2+4(B_1^2+B_2^2)},\\
p_2 &=& \frac{1}{2}-\frac{1}{2}\sqrt{B_3^2+4(B_1^2+B_2^2)}.
\end{eqnarray}
Therefore, the matrix $C$ has the following explicit form
\begin{widetext}
\begin{equation}
C~=~\frac{(p_1-p_2)^2}{4}\left(
\begin{array}{ccc}
2(\cos^2\xi\cos^2\psi+\sin^2\xi) & -\sin2\xi\sin^2\psi & -\cos\xi\sin2\psi\\
-\sin2\xi\sin^2\psi & 2(\cos^2\xi+\sin^2\xi\cos^2\psi) & -\sin2\psi\sin\xi\\
-\cos\xi\sin2\psi & -\sin2\psi\sin\xi & 2\sin^2\psi
\end{array}
\right).
\end{equation}
\end{widetext}

A straightforward calculation shows that the eigenvector corresponds to the maximum eigenvalue of the matrix $C$ has a two-fold degeneracy. Using the Schmidt orthogonalization, we can obtain two optimal directions with the form
\begin{eqnarray}
\vec{n}_o^1 &=& (-\sin\xi,\cos\xi,0),\\
\vec{n}_o^2 &=& (\cos\psi\cos\xi,\cos\psi\sin\xi,-\sin\psi).
\end{eqnarray}
It can be easily checked that the so-obtained optimal directions are orthogonal to the Bloch vector $\vec{B}$.


\begin{thebibliography}{60}
\bibitem{Breuer.07.NULL}H.-P. Breuer and F. Petruccione, {\it The Theory of Open Quantum Systems}(Oxford University Press, Oxford, 2007).
\bibitem{Weiss.12.NULL}U. Weiss, {\it Quantum Dissipative Systems}(World Scientific, Singapore, 2012).
\bibitem{Lindblad.76.CMP}G. Lindblad, Commun. Math. Phys. {\bf48}, 119(1976).
\bibitem{Li.11.PRA}C.-F. Li, J.-S. Tang, Y.-L. Li, and G.-C. Guo, Phys. Rev. A {\bf83}, 064102(2011).
\bibitem{Cialdi.11.PRA}S. Cialdi, D. Brivio, E. Tesio, and M. G. A. Paris, Phys. Rev. A {\bf83}, 042308(2011).
\bibitem{Liu.11.NP}B.-H. Liu, L. Li, Y.-F. Huang, C.-F. Li, G.-C. Guo, E.-M. Laine, H.-P. Breuer, and J. Piilo, Nat. Phys {\bf7}, 931(2011).
\bibitem{Smirne.11.PRA}A. Smirne, D. Brivio, S. Cialdi, B. Vacchini, andM. G. A. Paris, Phys. Rev. A {\bf84}, 032112(2011).
\bibitem{Barreiro.11.N}J. T. Barreiro, M. Muller, P. Schindler, D. Nigg, T. Monz, M. Chwalla, M. Hennrich, C. F. Roos, P. Zoller, and R. Blatt, Nature {\bf470}, 486(2011).
\bibitem{Gessner.14.NP}M. Gessner, M. Ramm, T. Pruttivarasin, A. Buchleitner, H.-P. Breuer, and H. Haffner, Nat. Phys {\bf10}, 105(2014).
\bibitem{Tang.15.O}J.-S. Tang et. al., Optica {\bf2}, 1014(2015).
\bibitem{Wolf.08.PRL}M. M. Wolf, J. Eisert, T. S. Cubitt, and J. I. Cirac, Phys. Rev. Lett. {\bf101}, 150402(2008).
\bibitem{Breuer.09.PRL}H.-P. Breuer, E.-M. Laine, and J. Piilo, Phys. Rev. Lett. {\bf103}, 210401(2009).
\bibitem{Rivas.10.PRL}A. Rivas, S. F. Huelga, and M. B. Plenio, Phys. Rev. Lett. {\bf105}, 050403(2010).
\bibitem{Lu.10.PRA}X.-M. Lu, X. Wang, and C. P. Sun, Phys. Rev. A {\bf82}, 042103(2010).
\bibitem{Chruscinski.11.PRA}D. Chru\'{s}ci\'{n}ski, A. Kossakowski, and A. Rivas, Phys. Rev. A {\bf83}, 052128 (2011).
\bibitem{Luo.12.PRA}S. Luo, S. Fu, and H. Song, Phys. Rev. A {\bf86}, 044101(2012).
\bibitem{Chruscinski.14.PRL}D. Chru\'{s}ci\'{n}ski and S.Maniscalco, Phys. Rev. Lett. {\bf112}, 120404(2014).
\bibitem{Breuer.16.RMP}H.-P. Breuer, E.-M. Laine, J. Piilo, and B. Vacchini, Rev. Mod. Phys. {\bf88}, 021002(2016).
\bibitem{Segal.05.PRL}D. Segal and A. Nitzan, Phys. Rev. Lett. {\bf94}, 034301 (2005).
\bibitem{Ren.10.PRL}J. Ren, P. H\"{a}nggi, and B. Li, Phys. Rev. Lett. {\bf104}, 170601(2010).
\bibitem{Boudjada.14.JPCA}N. Boudjada and D. Segal, J. Phys. Chem. A {\bf118}, 11323(2014).
\bibitem{Wang.15.SR}C.Wang, J. Ren, and J. Cao, Sci. Rep. {\bf5}, 11787 (2015).
\bibitem{Liu.16.NULL}J. Liu, H. Xu, B. Li, and C. Wu, arXiv:1609.05598(2016).
\bibitem{Carrega.16.PRL}M. Carrega, P. Solinas, M. Sassetti, and U. Weiss, Phys. Rev. Lett. {\bf116}, 240403(2016).
\bibitem{Guarnieri.16.PRA}G. Guarnieri, C. Uchiyama, and B. Vacchini, Phys. Rev. A {\bf93}, 012118(2016).
\bibitem{Schmidt.16.PRA}R. Schmidt, S.Maniscalco, and T. Ala-Nissila, Phys. Rev. A {\bf94}, 010101(2016).
\bibitem{Solinas.13.PRB}P. Solinas, D. V. Averin, and J. P. Pekola, Phys. Rev. B {\bf87}, 060508(2013).
\bibitem{Schmidt.15.PRB}R. Schmidt, M. F. Carusela, J. P. Pekola, S. Suomela, and J. Ankerhold, Phys. Rev. B {\bf91}, 224303(2015).
\bibitem{Guarnieri.16.PRAa}G. Guarnieri, J. Nokkala, R. Schmidt, S. Maniscalco, and B. Vacchini, Phys. Rev. A {\bf94}, 062101(2016).
\bibitem{Haikka.12.PRA}P. Haikka, J. Goold, S. McEndoo, F. Plastina, and S. Maniscalco, Phys. Rev. A {\bf85}, 060101(2012).
\bibitem{He.11.PRA}Z. He, J. Zou, L. Li, and B. Shao, Phys. Rev. A {\bf83}, 012108(2011).
\bibitem{Hauke.16.NP}P. Hauke, M. Heyl, L. Tagliacozzo, and P. Zoller, Nat. Phys. {\bf12}, 778(2016).
\bibitem{Chruscinski.10.PRL}D. Chru\'{s}ci\'{n}ski and A. Kossakowski, Phys. Rev. Lett. {\bf104}, 070406(2010).
\bibitem{Gorini.76.JMP}V.Gorini, A. Kossakowski, and E. C. G. Sudarshan, J. Math. Phys. (N.Y.) {\bf17}, 821(1976).
\bibitem{Breuer.04.PRA}H.-P. Breuer, Phys. Rev. A {\bf70}, 012106(2004).
\bibitem{Zeng.11.PRA}H.-S. Zeng, N. Tang, Y.-P. Zheng, and G.-Y. Wang, Phys. Rev. A {\bf84}, 032118(2011).
\bibitem{Leggett.87.RMP}A. J. Leggett, S. Chakravarty, A. T. Dorsey, M. P. A. Fisher, A. Garg, and W. Zwerger, Rev. Mod. Phys. {\bf59}, 1(1987).
\bibitem{Merkli.13.JMC}M. Merkli, G. P. Berman, and R. Sayre, J. Math. Chem. {\bf51}, 890(2013).
\bibitem{Garg.85.JCP}A. Garg, J. N. Onuchic, and V. Ambegaokar, J. Chem. Phys. {\bf83}, 4491(1985).
\bibitem{Fukai.05.NULL}Y. Fukai, {\it The Metal-Hydrogen System}(Springer-Verlag, Berlin, 2005).
\bibitem{Anderson.72.PM}P. W. Anderson, B. I. Halperin, and C. M. Varma, Philos. Mag. {\bf25}, 1(1972).
\bibitem{Haikka.10.PS}P. Haikka, Phys. Scr. {\bf2010}, 014047(2010).
\bibitem{Clos.12.PRA}G. Clos and H.-P. Breuer, Phys. Rev. A {\bf86}, 012115(2012).
\bibitem{Hao.13.JPA}X. Hao, N.-H. Tong, and S. Zhu, J. Phys. A: Math. Theor. {\bf46}, 355302(2013).
\bibitem{Helstrom.76.NULL}C. W. Helstrom, {\it Quantum Detection and Estimation}(Academic, New York, 1976).
\bibitem{Giovannetti.06.PRL}V. Giovannetti, S. Lloyd, and L. Maccone, Phys. Rev. Lett. {\bf96}, 010401(2006).
\bibitem{You.07.PRE}W. L. You, Y. W. Li, and S. J. Gu, Phys. Rev. E {\bf76}, 022101(2007).
\bibitem{Venuti.07.PRL}L. C. Venuti and P. Zanardi, Phys. Rev. Lett. {\bf99}, 095701(2007).
\bibitem{Zanardi.07.PRL}P. Zanardi, P. Giorda, and M. Cozzini, Phys. Rev. Lett. {\bf99}, 100603(2007).
\bibitem{Banchi.14.PRE}L. Banchi, P. Giorda, and P. Zanardi, Phys. Rev. E {\bf89}, 022102(2014).
\bibitem{Marzolino.14.PRA}U. Marzolino and T. Prosen, Phys. Rev. A {\bf90}, 062130(2014).
\bibitem{Hyllus.12.PRA}P. Hyllus, W. Laskowski, R. Krischek, C. Schwemmer,
W. Wieczorek, H. Weinfurter, L. Pezz\'{e}, and A. Smerzi, Phys. Rev. A {\bf85}, 022321(2012).
\bibitem{Braunstein.94.PRL}S. L. Braunstein and C. M. Caves, Phys. Rev. Lett. {\bf72}, 3439(1994).
\bibitem{Hyllus.10.PRA}P. Hyllus, O. G\"{u}hne, and A. Smerzi, Phys. Rev. A {\bf82}, 012337(2010).
\bibitem{Suzuki.16.PRA}J. Suzuki, Phys. Rev. A {\bf94}, 042306(2016).
\bibitem{Fujiwara.01.PRA}A. Fujiwara, Phys. Rev. A {\bf63}, 042304(2001).
\bibitem{Koh.07.PRB}Y. K. Koh and D. G. Cahill, Phys. Rev. B {\bf76}, 075207(2007).
\bibitem{Lefevre.05.PSI}S. Lef\`{e}vre and S. Volz, Phys. Sci. Instrum. {\bf76}, 033701(2005).
\bibitem{Chavez.14.APLM}E. Ch\'{a}vez-\'{A}ngel and et al., Appl. Phys. Lett. Mater. {\bf2}, 012113(2014).
\bibitem{Reparaz.14.PSI}J. Reparaz and et al., Phys. Sci. Instrum. {\bf85}, 034901 (2014).
\bibitem{Makri.95.JCP}N. Makri and D. E. Makarov, J. Chem. Phys. {\bf102}, 4600(1995).
\bibitem{Goerlich.89.EL}R. G\"{o}rlich, M. Sassetti, and U. Weiss, Europhys. Lett. {\bf10}, 507(1989).
\end{thebibliography}

\end{document}